\documentclass[letterpaper]{article} 
\usepackage{aaai24}  
\usepackage{times}  
\usepackage{helvet}  
\usepackage{courier}  
\usepackage[hyphens]{url}  
\usepackage{graphicx} 
\usepackage{natbib}  
\usepackage{caption} 
\usepackage{algorithm}
\usepackage{algorithmic}

\usepackage{multirow}
\usepackage{amsmath}
\usepackage{booktabs}
\usepackage{tabularx}
\usepackage{enumitem}
\usepackage{framed}
\usepackage{xcolor}
\usepackage{newfloat}
\usepackage{listings}
\DeclareCaptionStyle{ruled}{labelfont=normalfont,labelsep=colon,strut=off} 
\floatstyle{ruled}
\newfloat{listing}{tb}{lst}{}
\floatname{listing}{Listing}
\title{``Hi. I'm Molly, Your Virtual Interviewer!'' --- Exploring the Impact of Race and Gender in AI-powered Virtual Interview Experiences}
\author{
    Shreyan Biswas\textsuperscript{\rm 1}\equalcontrib,
    Ji-Youn Jung\textsuperscript{\rm 2}\equalcontrib, 
    Abhishek Unnam\textsuperscript{\rm 3} \\
    Kuldeep Yadav\textsuperscript{\rm 3}
    Shreyansh Gupta\textsuperscript{\rm 3}
    Ujwal Gadiraju\textsuperscript{\rm 1}
}
\affiliations{
    \textsuperscript{\rm 1} Delft University of Technology\\
    \textsuperscript{\rm 2} Carnegie Mellon University\\
    \textsuperscript{\rm 3} SHL Labs\\
    s.biswas@tudelft.nl, jiyounj@andrew.cmu.edu, Abhishek.Unnam@shl.com, \\kuldeep.yadav@shl.com, shreyansh.gupta@shl.com, u.k.gadiraju@tudelft.nl
}

\begin{document}
\maketitle

\newcommand{\noagency}[1]{\textbf{\texttt{No Agency}}}
\newcommand{\agency}[1]{\textbf{\texttt{Agency}}}
\newcommand{\female}[1]{\textbf{\texttt{Female}}}
\newcommand{\male}[1]{\textbf{\texttt{Male}}}
\newcommand{\nonb}[1]{\textbf{\texttt{Non-binary}}}
\newcommand{\black}[1]{\textbf{\texttt{Black}}}
\newcommand{\white}[1]{\textbf{\texttt{White}}}

\newcommand{\ujwal}[1]{\textcolor{cyan}{[\textbf{UG: #1}]}}
\newcommand{\ji}[1]{\textcolor{magenta}{[\textbf{JI: #1}]}}
\newcommand{\sh}[1]{\textcolor{orange}{[\textbf{SH: #1}]}}

\begin{abstract}
The persistent issue of human bias in recruitment processes poses a formidable challenge to achieving equitable hiring practices, particularly when influenced by demographic characteristics such as gender and race of both interviewers and candidates. Asynchronous Video Interviews (AVIs), powered by Artificial Intelligence (AI), have emerged as innovative tools aimed at streamlining the application screening process while potentially mitigating the impact of such biases. These AI-driven platforms present an opportunity to customize the demographic features of virtual interviewers to align with diverse applicant preferences, promising a more objective and fair evaluation. Despite their growing adoption, the implications of virtual interviewer identities on candidate experiences within AVIs remain under explored. We aim to address this research and empirical gap in this paper. To this end, we carried out a comprehensive between-subjects study involving 218 participants across six distinct experimental conditions, manipulating the gender and skin color of an AI virtual interviewer agent. Our empirical analysis revealed that while the demographic attributes of the agents did not significantly influence the overall experience of interviewees, variations in the interviewees' demographics significantly altered their perception of the AVI process. Further, we uncovered that the mediating roles of Social Presence and Perception of the virtual interviewer critically affect interviewees' \textit{Perceptions of Fairness} (+), \textit{Privacy} (-), and \textit{Impression management} (+). 


\end{abstract}

\section{Introduction}

The evolution of personnel selection interviews has been profound, with research tracing back over a century \cite{Moore_1921}. This wide spectrum of scholarly work has delved into the intricate social dynamics of interviews \cite{Fletcher_1992, McCarthy_Goffin_2004} and has increasingly sought to leverage technological advancements to enhance the efficiency and scalability of the interview processes \cite{Blacksmith_Willford_Behrend_2016}. The COVID-19 pandemic has catalyzed the adoption of innovative interviewing techniques, with a notable shift towards Asynchronous Video Interviews (AVI) powered by AI. This shift is evidenced by the widespread adoption of automated screening tools among Fortune 500 companies \cite{hanson2023forbes}, highlighting the pivotal role of AVIs in modern recruitment strategies \cite{Dunlop_Holtrop_Wee_2022}.

Prior research has shown that job applicants experience \textit{stereotype threat} during the job selection process~\cite{graves2008sex, Whysall_2018}, a psychological state where individuals may underperform due to anxieties about confirming negative stereotypes associated with their racial, ethnic, gender, or cultural identities~\cite{steele1995stereotype}. Studies have documented the impact of demographic congruence between interviewers and interviewees on applicant perceptions and strategies highlighting the intricate ways in which demographic factors shape interview outcomes \cite{pedulla2014positive, jaquemet2012indiscriminate, landy2008stereotypes, opie2015hair, latu2015gender, goldberg2003applicant, previtali2023ageism}.

Despite the flexibility and scalability of AVIs, concerns around perceived fairness, trust, privacy, diminishing perception of social presence, and reduced capabilities to utilize impression management tactics continue to persist \cite{hbrWhereAutomated, Basch2021-go, roulin2023bias, Langer_König_Krause_2017, liu2023speech, Blacksmith_Willford_Behrend_2016}.
Although efforts have been made to address these challenges through algorithmic fairness \cite{fabris2023fairness}, providing explanations \cite{Basch_Melchers_2019}, and implementing post-hoc measures \cite{raghavan2020mitigating}, the exploration of interviewee demographics and their nuanced perceptions during the AVI process remains a relatively underexplored avenue. This understanding can play a pivotal role in designing an equitable interview experience across diverse candidate pools.

Our study investigates how variations in the gender and race of an AVI agent affect interviewees' experiences, aiming to distil insights that can inform the future design of AVI agents, addressing the following research questions: 

\noindent
\textbf{RQ1:} How do the gender and race of a virtual interviewer impact an interviewee’s virtual interview experience?

\noindent
\textbf{RQ2:} How do an interviewee's gender and race influence their virtual interview experience? 

We conducted a 3$\times$2 between-subject study simulating the AVI screening phase using an agent with varied gender (female, male, non-binary) and racial (black, white) configurations across conditions. Recruiting a diverse participant pool, we 
shed light on the impact of AVI agent configurations on interviewee experiences, focusing on \textit{perceived fairness}, \textit{social perception and presence}, \textit{privacy and emotional response}, and \textit{impression management} tactics. 
\section{Background and Related Work}

\textbf{Asynchronous Video Interviews.} 
Traditionally segmented into sourcing, screening, interviewing, and candidate selection phases \cite{bogen2018help}, the screening stage has evolved with the introduction of Asynchronous Video Interviews (AVIs). A derivative of technology-mediated interviews (TMIs), AVIs offer a scalable solution to assess candidates beyond their resumes through pre-recorded video responses \cite{Brenner_Ortner_Fay_2016, kleinlogel_schmid_et_al}. This method promises standardization and fairness by providing all candidates with identical questions, eliminating the variability inherent in live interactions \cite{Moore_Kearsley_1996, Rasipuram_Rao_Jayagopi_2016}.
Despite these advantages, AVIs face criticism for lacking interactivity, raising privacy concerns, diminishing social presence, and limiting non-verbal communication cues crucial for impression management \cite{roulin2023bias, liu2023speech, Blacksmith_Willford_Behrend_2016}. Research efforts have thus shifted towards enhancing the AVI experience from the interviewee's perspective, exploring features like re-recording options and explanatory feedback to improve perceptions of trust and fairness \cite{Roulin_Wong_Langer_Bourdage_2022, Basch_Melchers_2019}.

\textbf{Designing AVI Agents.}

\citet{li2017confiding} utilized text-based conversational agents (CAs) with varying personalities and found that a virtual interviewer can make a recruiting process more efficient,  objective,  and inclusive. In the context of AVIs, prior work highlighted the potential of embodied CAs to generate dynamic follow-up questions, thereby tackling the monotony of AVI dialogues and fostering a more authentic, human-like interview experience~\cite{RaoS_Bimproving}. Further investigation by ~\citet{Thakkar_Rao_Shubham_Jain_Jayagopi_2022} into the impact of verbal and nonverbal cues from virtual interviewers corroborated the positive influence on interviewees' experiences. However, these studies have yet to address the impact of demographic features (i.e., gender and race) of interviewer agents on interviewee experiences within the AVI process.

\textbf{Fairness in Algorithmic Hiring---The Interviewee Standpoint.} Previous research has investigated fairness in algorithmic hiring.
Efforts have been made to reduce socio-linguistic bias in resumes~\cite{deshpande2020mitigating}, and studies have surveyed the potential benefits of algorithmic hiring~\cite{fabris2023fairness}. 
Real-world applications of algorithmic pre-employment assessments have also been reported~\cite{raghavan2020mitigating}. However, there is a lack of research focusing on fairness from the applicants' perspective, especially regarding stereotype threats and perceived similarities during interviews. \textit{Stereotype threat}, a cognitive bias in personnel selection, suggests that certain environments can perpetuate stereotypes of specific groups being less competent, potentially impairing performance~\cite{steele1995stereotype, schmader2010stereotype}. The Stereotype Content Model (SCM), explaining interpersonal impressions along dimensions of perceived warmth and competence, is also relevant here~\cite{fiske2018model}. SCM has been used in HCI research to understand the social aspects of technology, and we adopt this model in our study to gauge interviewees' perceptions in different AVI scenarios.

Prior research shows that demographic similarities between interviewers and interviewees (e.g., gender, race, and age) can influence hiring decisions and the interviewee's strategy for rapport-building~\cite{landy2008stereotypes, opie2015hair, harrington2023trust, previtali2023ageism, atkins1988recruiters, francesco1981gender}. These findings underline the importance of considering demographic factors in the design of AVIs to foster an inclusive and equitable interviewing environment.

Our study contributes to this by exploring the effects of demographic characteristics of AI interviewers on the perceptions and experiences of interviewees in AVI screening interviews. 
Our work extends beyond the interviewer's attributes to also examine how interviewees from diverse backgrounds perceive the AVI process, unveiling the intricate interplay between their demographic characteristics and the subjective experience in interviews with virtual interviewer agents.

\section{System Design}

\subsection{Screening Interview Scenario}
\label{section:interview_flow}

\textbf{Interview Flow.}

We 
targeted the screening phase of recruitment due to 
The prevalent use of AVIs during this initial stage is when HR personnel commonly conduct brief phone interviews before progressing candidates to the main interview process. 
The interview process consisted of three stages:
(1) An \textit{introduction} comprising standard introductory inquiries; 
(2) a \textit{behavioral} phase with with hypothetical work scenario questions; and
(3) a \textit{screening} phase with questions to evaluate candidates' qualifications, availability, and salary expectations, with an opportunity for candidates to ask clarifying questions. Each stage included predefined questions and follow-ups as needed. A full list of questions is in the supplementary material. 
For our controlled experiment, we used a task-oriented dialogue setting on our AVI platform, allowing automatic turn-taking and flexible conversations. 
In this study, participants completed a screening interview process for a `Customer Support Specialist' role, such as retail salespersons, cashiers, and customer service representatives, 
given that these professions represent some of the largest occupational groups in the USA 
\cite{USjobstatistics2023}.

\noindent\textbf{Design of the Agent Avatars.}
 
\begin{figure}[t]
    \centering
      \includegraphics[width=0.6\columnwidth, keepaspectratio]{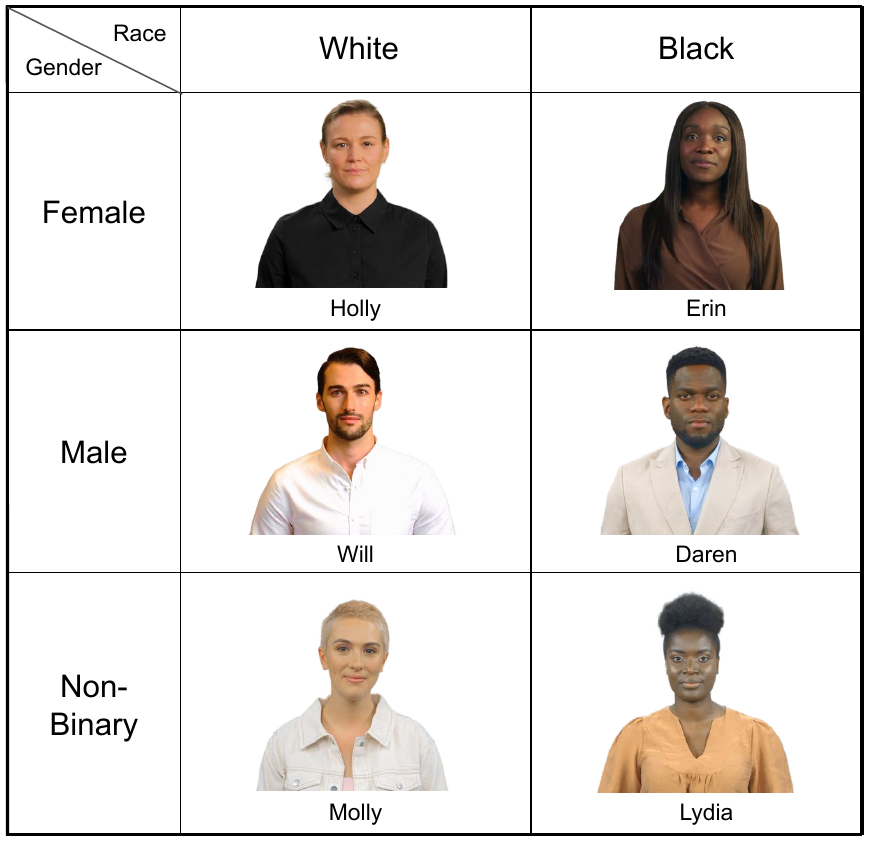}
    \caption{Avatars of the virtual interviewer agent.}
    \label{fig:avatars}
\end{figure}

Prior work has demonstrated that individuals perceive conversational agent avatars as more `professional' when they exhibit greater realism~\cite{ring2014right}. Consistent with common practice in existing AVI tools, we employed highly realistic human-like avatars.
Figure~\ref{fig:avatars} shows the 
avatars that were used across different \textit{genders} (female, male, and non-binary) and \textit{race} (white vs. black) conditions. 
It is important to note that non-binary identity is self-defined and can vary widely; thus, our avatars were crafted to be inclusive without necessarily adhering to strict gender androgyny. Despite the different avatars, every agent behaved consistently in the same way, speaking the same dialogue with the same accent, nodding, smiling, and making eye contact.

\subsection{System Development}
We developed a web-based AVI platform for virtual interviews. The system consists of the following components:

\noindent\textbf{Voice Activity Detection (VAD) and Speech to Text.}

To facilitate automatic turn-taking during interviews, we implemented a real-time VAD based on~\citet{7077834}. This component utilizes an adaptive silence-duration threshold to discern speech cessation, optimizing for varied response lengths inherent in different question types. Longer thresholds account for situational-behavioral questions and shorter ones for concise screening questions. 

Transcription is performed using the Microsoft Azure Speech Recognition engine.

\noindent\textbf{Conversation Understanding and Management.}

The two following modules manage the flow of the conversation: 

\begin{itemize}[nosep, leftmargin=*]
    \item \textit{Spoken Language Understanding (SLU)}: Mapping transcribed text to over 45 predefined intents. 
    \item \textit{Dialogue Management}: Essential for structured interviewing, this module ensures all requisite questions are posed, updating the conversation state after each exchange to determine subsequent actions/responses based on 
    intents.
\end{itemize}

We integrated Google Dialogflow\footnote{\url{https://cloud.google.com/dialogflow/}} to utilize its advanced intent classifier and dialogue management capabilities customizing it with our dataset to refine conversational flows.
\noindent\textbf{Video Dialogue Response Selection.}

Consistent with \citet{rizi2023does}, which highlights the benefits of visual feedback on social presence and interviewee performance, our system uses the Synthesia\footnote{\url{https://www.synthesia.io/}} platform to create and store human-like avatar video snippets for each dialogue.

\noindent\textbf{Intelligent Video Player.}
This component assembles video snippets based on the dialogue cues to form a cohesive narration. To address potential latency issues~\cite{peng2020understanding}, it includes conversational fillers and non-verbal cues (e.g., nodding, eye-blinking).

\noindent\textbf{External Data Sources.}
Our conversational engine uses a knowledge base and API to enrich context with candidate details, job specifics, and interview questions. An AI model generates relevant follow-up questions, creating a dynamic and interactive interview experience.

\section{Method}

We conducted a randomized 3 (gender) $\times$ 2 (race) factorial between-subjects study to explore the impact of an AVI agent's and the interviewee's demographic attributes on participants' \textit{perceived fairness}, \textit{social perception and presence}, \textit{privacy and emotional response}, and \textit{impression management} tactics. Participants were recruited from the online crowdsourcing platform Prolific and subsequently assigned randomly to one of six virtual interviewer conditions: \texttt{Black-Female}, \texttt{Black-Male}, \texttt{Black-Non Binary}, \texttt{White-Female}, \texttt{White-Male}, and \texttt{White-Non Binary}. Following this assignment, they engaged in the interview process.

This study received institutional ethics approval.

\subsection{Participants}

A G-Power analysis~\cite{faul2007g} determined that a sample size of 211 participants was required (\textit{f=0.25}; $\alpha$\textit{=0.05, 1-}$\beta$\textit{=0.8}). To accommodate potential exclusions, we recruited 236 workers. 
Participation was restricted to individuals from the US or UK with customer-facing job experience, a minimum approval rating of 95\%, and consent to video recording during a simulated interview. Twelve participants were excluded for failing an attention check, and six non-binary participants were excluded to address sample skewness.\footnote{Note that only the data from participants were excluded from the analysis, their compensation was not withheld.} 

The remaining sample (\textit{N=218}; 60 white males, 60 white females, 55 black males, 43 black females) had a mean age of 35.4 years (\textit{SD = 8.75}), with ages ranging from 19 to 66. Participants were compensated at an hourly rate of USD \$15, and measures were taken to prevent repeated participation.

\subsection{Measures}~\label{section:measures}

Our measure of overall AVI experience is comprised of outstanding variables that were often explored from the previous literature. Table~\ref{table:list_of_variables} shows the complete summary of independent variables, dependent variables, and covariates that we measured through pre- and post-study surveys.

\begin{table*}[t]
\centering
\small
\begin{tabular}{p{3.5cm} p{3.5cm} p{9cm}}
\toprule
\textbf{Variable Types} & \textbf{Variable Names} & \textbf{Description} \\ \midrule
\textit{\textbf{Independent Variable}} & Interviewer Demographic & (i) White-Male, (ii) White-Female, (iii) White-Non Binary, (iv) Black-Male, (v) Black-Female, (vi) Black-Non Binary \\ \cmidrule{2-3} 
& Participant Demographic & (i) Black-Female, (ii) White-Female, (iii) Black-Male, (iv) White-Male \\ \midrule
\multirow{3}{*}{\textit{\textbf{Dependent Variables}}} & PF (10 Qs) & Procedural (3 Qs): Fairness in interview process and methods. \\ \cmidrule{3-3}
& & Behavioral (3 Qs): Fairness in the organization's hiring decision. \\ \cmidrule{3-3}
& & Interactional (4 Qs): Fairness in respectfulness and informativeness of interview. \\ \cmidrule{2-3} 
& SPP (8 Qs) & Perceived warmth and competence (4 Qs): Evaluation of friendliness and competence of interviewer. \\ \cmidrule{3-3}
& & Social presence (4 Qs): Comfort and perception of interviewer's social role. \\ \cmidrule{2-3} 
& PER (10 Qs) & Privacy concerns (5 Qs): Apprehension about privacy, risk of invasion, and data misuse. \\ \cmidrule{3-3}
& & Emotional creepiness (5 Qs): Feelings of discomfort, unease, and fear during the interview. \\ \cmidrule{2-3} 
& IM (4 Qs) & Effectiveness in presenting skills, knowledge, and nonverbal cues during interview. \\ \midrule
\textit{\textbf{Covariates}} &  ATI (9 Qs) & Interest and engagement in using and exploring technology. \\ \cmidrule{2-3} 
& MASI (30 Qs) & Anxiety in interviews: communication, appearance, discomfort, stress, and physical symptoms. \\ \bottomrule
\end{tabular}
\caption{A list of dependent, independent variables, and covariates considered in our study.}
\label{table:list_of_variables}
\end{table*}

\noindent\textbf{Dependent variables.}
\label{subsec:dv}
\textit{Perceived Fairness (PF)} 
was measured using a 7-point Likert scale with questions adopted from previous studies~\cite{norskov2020applicant, bauer2001applicant, mclarty2016dispositional}. This scale captures domain-specific fairness through three dimensions of the AVI agent's procedural, interactional fairness, and behavioral intentions.

\textit{Social Perception and Presence (SPP)}
was defined as a compound measure that captures the interviewees' stereotypical perceptions alongside their sense of the AVI agent's social presence. 
We employ a 7-point Likert scale across four questions based on~\cite{jung2022great, halkias2020universal} to evaluate the interviewees' views on the interviewer's perceived warmth and competence. Additionally, to quantify social presence — reflecting the degree of social interaction within the interview setting — we incorporate four additional questions, also rated on a 7-point Likert scale. Social presence captures the social interaction aspect of an interview and has been used in the past as a key metric in designing social robots/virtual AI agents \cite{liu2023speech, lee2006can}.

\textit{Privacy and Emotional Response (PER)}: Asynchronous interviews have been perceived as privacy-intrusive and creepy~\cite{roulin2023bias, Langer_König_Krause_2017}. We measured the perceived emotional creepiness and privacy concerns 
through ten questions on a 7-point Likert scale, adopted from
\citet{Langer_König_Krause_2017}. 

In the domain of personnel selection interviews, \textit{Impression Management (IM)} has often been touted as a key behavioral tactic that interviewees use to build rapport with interviewer \cite{Fletcher_1992, Roulin_Pham_Bourdage_2023} and has also been shown to affect interviewee performance in an interview \cite{Roulin_Wong_Langer_Bourdage_2022}. This scale was adopted from~\cite{Basch2021-go, tsai2005exploring} and contains 4 questions with a 5-point Likert scale.

\noindent\textbf{Covariates.}
\label{subsec:cov}
\textit{Affinity to Technology (ATI)}: 
Research has shown how people's affinity for technology impacts their judgment towards a new technology~\cite{franke2019personal}, and their perceived trust in an intelligent system~\cite{tolmeijer2021second}. 
We adopted and administered the validated 9-item `Affinity for Technology Interaction (ATI)' scale, which participants completed before the main interview session. 

\textit{Measure of Anxiety in Selection Interviews (MASI)}: We used a validated 
scale to measure an individual's perceived stress in the interviewing environment~\cite{mccarthy2004measuring}. MASI consists of five dimensions: Communication, Appearance, Social, Performance, and Behavioral anxiety. It was measured through a set of 30 questions consisting of a 5-point Likert scale. This measurement was captured post-interview to not influence the psychological state of participants during the interview.

\subsection{Experimental Setup and Procedure}

Participants entered our experiment via an interview platform, where they were informed about the task and gave consent for video and audio recording. They completed pre-task questionnaires to provide demographic information (gender and ethnicity), measure their affinity for technology interaction (ATI scale), and assess their anxiety in selection interviews (MASI scale). An attention question was also asked to ensure the quality of the response.
Participants were randomly assigned to interview with one of the six virtual interviewer agents; 3 \texttt{genders} (i.e., \female{}, \male{}, and \nonb{}) X 2 \texttt{race} (i.e., \black{} and \white{}). Information about the virtual interviewers' age, role, and gender was standardized and displayed with their images. During the live interview, participants interacted with the virtual interviewer and completed a post-task survey afterwards.

\noindent\textbf{Statistical Analysis.}

We first used a one-way ANCOVA to examine how interviewer demographics affect interviewee perceptions. Next, we explored how interviewee demographics influenced their perceptions of the interviewer and identified any significant differences across groups. 
We then conducted a mediation analysis using the Pingouin library in Python~\cite{Vallat2018} to explore how interviewee's perceptions of the AVI agent's Social Presence and Perception (SPP) mediated key outcomes: Perceived Fairness (PF), Privacy and Emotional Response (PER), and Impression Management (IM). For this analysis, we converted user demographic data into binary format, with 1 indicating a specific demographic attribute and 0 for others. We adopted the mediation framework by \citet{Baron_Kenny_1986}, incorporating a bias-corrected, non-parametric bootstrap approach \cite{Efron_1987} to estimate the indirect effect. All the analyses were adjusted for the covariates and each analysis was conducted with all sets of dependent variables. 
For all ANCOVA analyses, we validated the normality assumption (Shapiro-Wilk test) and homogeneity of variance (Levene's test). If any of the assumptions were violated, we performed the non-parametric Kruskal-Wallis test and followed it up with a Games-Howell post hoc test which does not require the assumptions of normality or homogeneity of variance to hold. If any significant ANCOVA findings were discovered then they were further explored through standard post-hoc parametric t-tests with Tukey-HSD adjustments.

\section{Results}

\noindent\textbf{Main Effect of AVI Agent Demographics.}
\label{subsubsec:results_id}

Through a Kruskal-Wallis test, we found no significant differences for Perceived Fairness ($\chi^2$ = 1.390, p = .92, df = 5), Social Presence and Perception ($\chi^2$ = 2.095, p = .83, df = 5), Privacy and Emotional Response ($\chi^2$ = 2.050, p = .84, df = 5), Impression Management ($\chi^2$ = 1.722, p = .88, df = 5), and Perceived Outcome ($\chi^2$ = 1.915, p = .86, df = 5)

\noindent\textbf{Main Effect of Participant Demographics.}
\label{subsubsec:results_pd}

A Kruskal-Wallis test showed significant differences in Perceived Fairness \(\chi^2 = 28.23\), \(p < .001\), indicating notable demographic effects. Post-hoc Games-Howell tests revealed significant differences among groups: black females (\(M = 5.94\), \(SD = 0.78\)) vs. white females (\(M = 5.28\), \(SD = 1.0\)), \(t(100.24) = 3.71\), \(p = .002\); white females vs. black males (\(M = 6.09\), \(SD = 1.05\)), \(t(110.86) = -4.23\), \(p < .001\); and black males vs. white males (\(M = 5.50\), \(SD = 0.98\)), \(t(110.19) = 3.15\), \(p = .01\).

We also found significant participant demographic influences on Social Presence and Perception \(\chi^2 = 34.25\), \(p < .001\). Differences were noted between black females (\(M = 5.39\), \(SD = 1.17\)) vs. white females (\(M = 4.67\), \(SD = 1.18\)), \(t(90.88) = 3.06\), \(p = .01\); white females vs. black males (\(M = 5.7\), \(SD = 1.4\)), \(t(105.85) = -4.273\), \(p < .001\); and black males vs. white males (\(M = 4.79\), \(SD = 1.14\)), \(t(104.24) = 3.81\), \(p = .001\).

Conversely, participant demographics did not influence Privacy and Emotional Response \(\chi^2 = 0.78\), \(p = .85\). However, Impression Management, \(\chi^2 = 17.69\), \(p < .001\) was significantly affected by participant demographics. The significant contrasts were noted between black females (\(M = 4.09\), \(SD = 0.66\)) vs. white females (\(M = 3.69\), \(SD = 0.82\)), \(t(99.60) = 2.77\), \(p = .03\); black females vs. white males (\(M = 3.68\), \(SD = 0.83\)), \(t(99.86) = 2.78\), \(p = .03\); white females vs. black males (\(M = 4.14\), \(SD = 0.84\)), \(t(111.58) = -2.94\), \(p = .02\); and black males vs. white males \(t(111.84) = 2.95\), \(p = .02\).

\begin{framed}
   \noindent
    \textbf{\textit{Summary}}: 
    {We found a statistically significant difference between participants' demographic factors on Perceived Fairness, Social Presence and Perception, and Impression Management. Overall, black male and black female participants reported higher scores in Perceived Fairness, Social Presence and Perception, and Impression Management in the overall AVI process, compared to the white male and white female participants. However, there were no significant differences between the perception of Privacy and Emotional Response.}
\end{framed}

\begin{figure*}[ht]
    \centering
    \includegraphics[width=\textwidth, keepaspectratio]{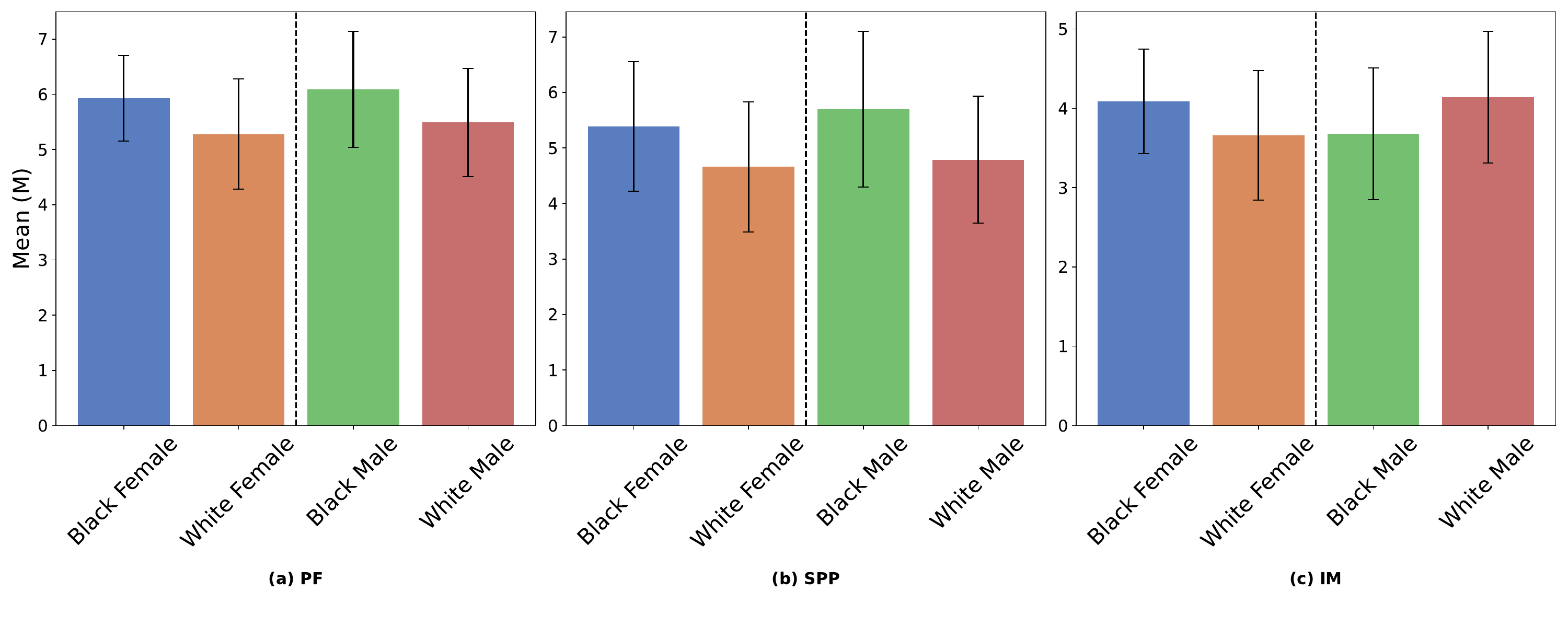}
    \caption{Comparison of statistically significant group-wise differences of participant demographics (Black Female vs. White Female etc.) on key interview metrics: (a) Perceived Fairness (PF), (b) Social Presence and Perception (SPP), and (c) Impression Management (IM). Subfigures (a, b, c) display mean scores for different demographic groups, with error bars representing standard deviations.}
    \label{fig:u_demo_direct_sig}
\end{figure*}

\subsection{Mediation Effect of Participant Demographics on AVI Agent Perceptions}

{Research has emphasized the significance of an AI agent's perceived social presence in establishing trust~\cite{gefen2004consumer}. Studies indicate that a heightened social presence can alleviate discomfort or ``creepiness" in human-AI interactions \cite{Oh_Bailenson_Welch_2018, lukacik2022into}. However, increased perceived humanness may also trigger eeriness due to the uncanny valley effect~\cite{Thaler_Schlogl_Groth_2020}, potentially negatively impacting the interview experiences. 
Therefore, we investigated how an AVI agent's Social Presence and Perception (SPP) mediates interviewees' perceptions, focusing on Fairness (PF), Privacy (PER), and Impression Management (IM) tactics. 
{This contrasts with earlier analyses of interviewees' perceptions of the AI avatar's demographic features (e.g., race and gender). This experiment shifts the focus to the socially constructed attribute (SPP) of the AI avatars.}}

\subsection{Black Female Participants}
For black female participants, our analysis did not indicate a significant impact of their identity on their social perception and presence (SPP) ratings of the AI interviewer (\textit{B = 0.376, 95\% CI [-0.05, 0.08], p = .088}).

\noindent\textbf{Mediation Effect on Perceived Fairness.}
Although we observed a significant total effect on Perceived Fairness rating  (\textit{B = 0.351, 95\% CI [-0.01, 0.68], p = .041}), neither the direct nor the indirect effects mediated through SPP were significant (\(p_{\text{direct}} = .26\); \(p_{\text{indirect}} = .07\)), suggesting no mediation effect of SPP on fairness perception.

\noindent\textbf{Mediation Effect on Privacy and Emotional Response.}
No significant effects—total, direct, or indirect—were observed on privacy concerns rating through SPP (\(p_{\text{total}} = .63\); \(p_{\text{direct}} = .82\); \(p_{\text{indirect}} = .07\)).

\noindent\textbf{Effect on Impression Management}
For black females, the role of SPP as a mediator on impression management (IM) score was not significant, with a direct effect of (\textit{B = 0.1, 95\% CI [-0.11, 0.31], p = .34}), and an indirect effect of (\textit{B = 0.151, 95\% CI [-0.004, 0.30], p = .072)}.

\subsection{White Female Participants}
Identifying as a white female correlated with a reduced perception of SPP compared to other demographics  (\textit{B = -0.547, 95\% CI [-0.928, -0.167], p = .005}).

\noindent\textbf{Mediation Effect on Perceived Fairness.}
The total effect on Perceived Fairness was significant (\textit{B = -0.46, 95\% CI [-0.757, -0.163], p = .003}), primarily due to a significant indirect effect through SPP (\textit{B = -0.340, 95\% CI [-0.586, -0.150], $p<.001$}), while the direct effect was not significant (\(p_{\text{direct}} = .19\)). This finding indicates that social presence and perception scores mediate fairness perceptions among White female interviewees.

\noindent\textbf{Mediation Effect on Privacy and Emotional Response.}
The direct effect on privacy concerns was significant (\textit{B = -0.417, 95\% CI [-0.73, -0.10], p = .009}), as was the indirect effect through SPP (\textit{B = 0.210, 95\% CI [0.08, 0.37], $p<.001$}); but the total effect was not significant ($p_{\text{total}}=.23$).

\noindent\textbf{Effect on Impression Management.}
A significant mediation was observed in the relationship between white female identity and their utilization of impression management tactics (IM), highlighted by a notable indirect effect through SPP (\textit{B = -0.226, 95\% CI [-0.38, -0.08], $p< .001$}), with both the direct and total effects being insignificant (\(p_{\text{direct}} = .36\); \(p_{\text{total}} = .25\)).

\subsection{Black Male Participants}
For black male participants, their identity was correlated with a heightened perception of Social Perception and Presence (SPP) of the AVI agent (\textit{B = 0.764, 95\% CI [0.37, 1.15], $p<.001$}) as compared to other demographics.

\noindent\textbf{Mediation Effect on Perceived Fairness.}
A strong mediation effect of SPP on perceived fairness was evident, with a significant total effect (\textit{B = 0.487, 95\% CI [0.18, 0.79], p = .002}) and a significant indirect effect through SPP (\textit{B = 0.480, 95\% CI [0.21, 0.76], $p<.001$}), although the direct effect was insignificant (\(p_{\text{direct}} = .94\)). 

\noindent\textbf{Mediation Effect on Privacy and Emotional Response.}
A significant direct effect on privacy and emotional response was identified (\textit{B = 0.583, 95\% CI [0.26, 0.91], $p<.001$}), along with a noteworthy negative indirect effect through SPP (\textit{B = -0.310, 95\% CI [-0.56, -0.14], $p < .001$}), despite the total effect being not significant (\(p_{\text{total}} = .12\)).

\noindent\textbf{Effect on Impression Management}
A substantial mediation of SPP on IM strategies was observed among Black male participants, with a significant indirect effect (\textit{B = 0.317, 95\% CI [0.14, 0.54], $p < .001$}).

\begin{figure}[ht]
    \centering
    \includegraphics[width=.8\columnwidth, keepaspectratio]{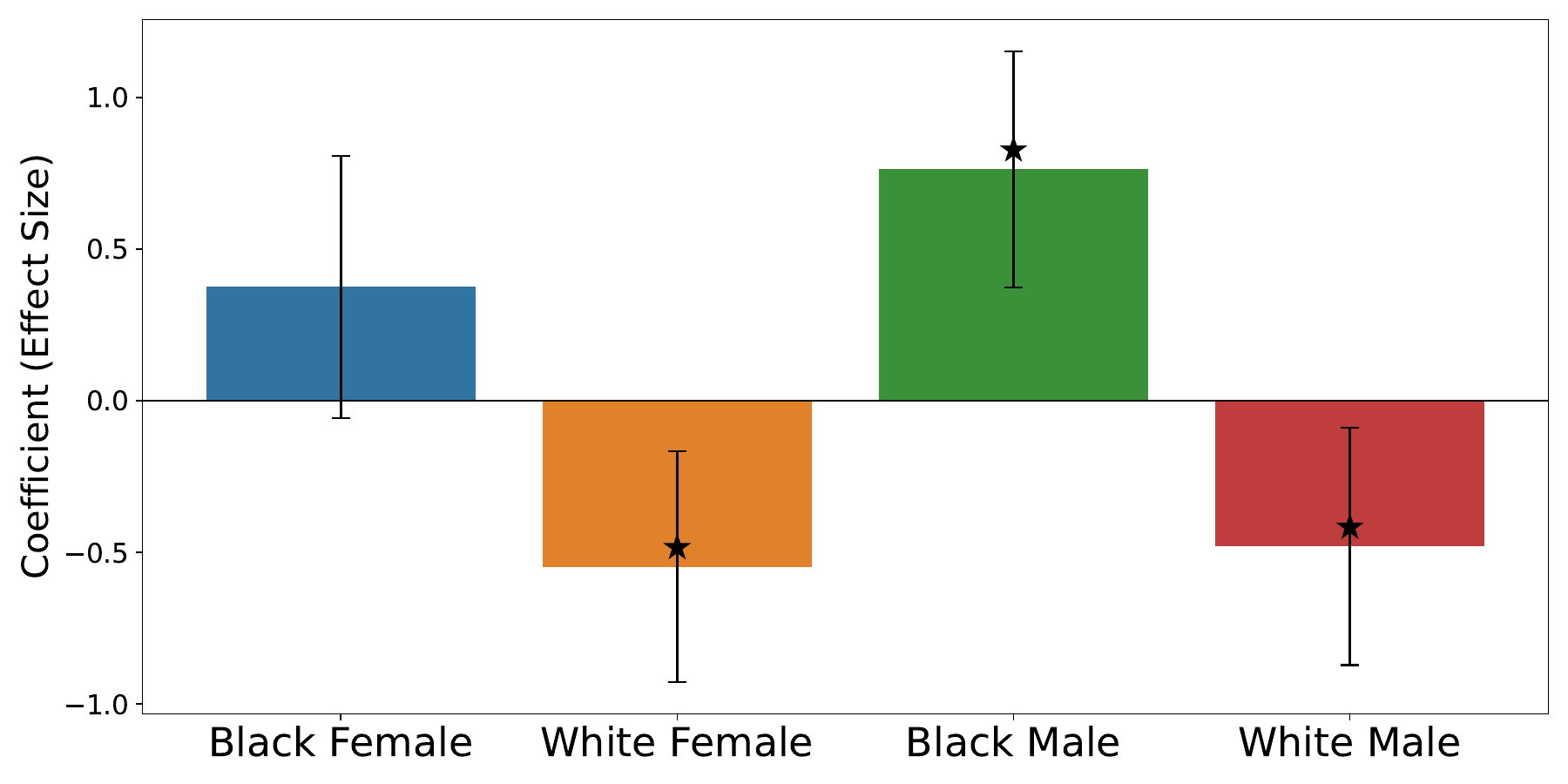}
    \caption{Effect of Participant Demographics on Social Presence and Perception (SPP): Coefficients represent the magnitude and direction of the impact each demographic has on SPP, highlighting how perceptions of social presence and perception vary across different groups. Each bar represents the effect size with corresponding 95\% confidence intervals depicted through error bars. Significant findings are highlighted with a star($\star$).}
    \label{fig:spp_pd}
\end{figure}

\begin{figure*}[ht]
    \centering
    \includegraphics[width=.8\textwidth, keepaspectratio]{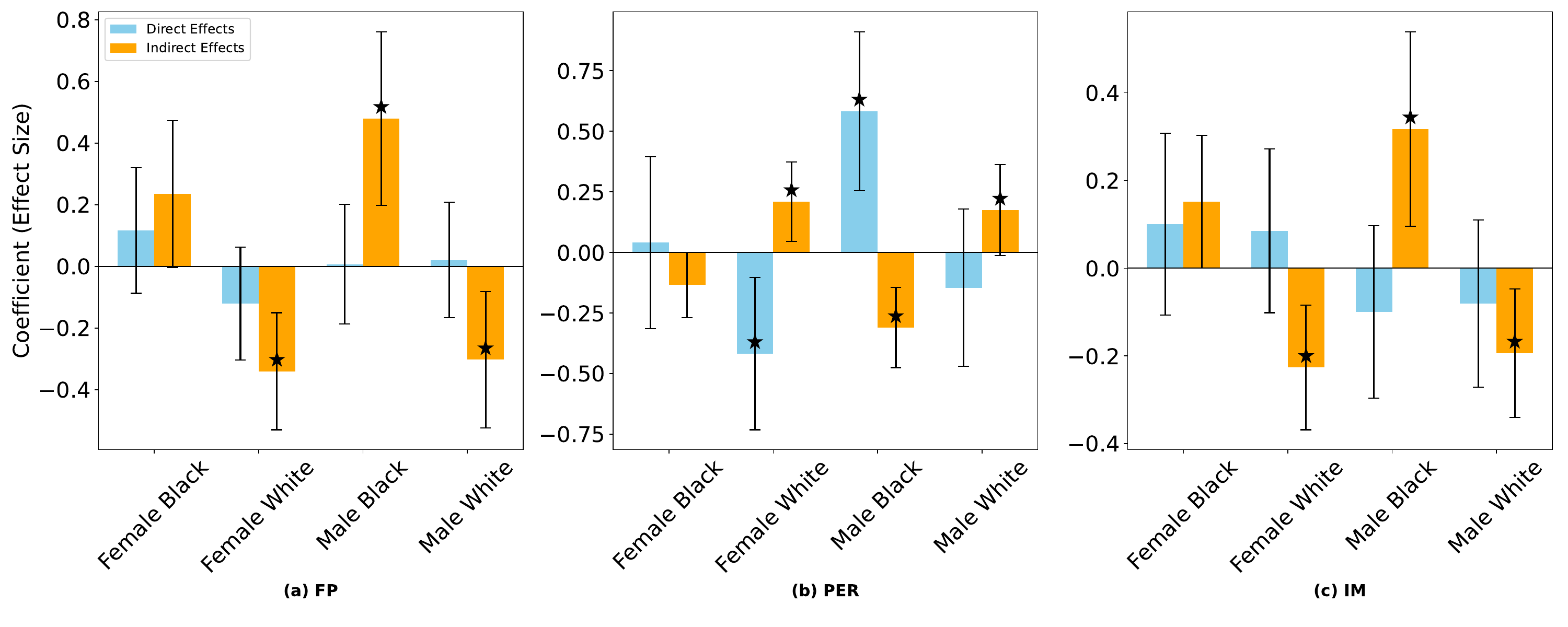}
    \caption{Direct and Indirect effects across user demographics under the mediation of Social Presence and Perception (SPP). Subplots (a), (b), and (c) illustrate the effects on Fairness Perception (FP), Privacy and Emotional Response (PER), and Impression Management (IM), respectively. Each bar represents the effect size with corresponding 95\% confidence intervals depicted through error bars. Significant findings are highlighted with a star($\star$).}

    \label{fig:direct_indirect}
\end{figure*}

\subsection{White Male Participants}
Identifying as a white male participant correlated with a reduced perception of SPP of the AVI agent compared to other demographics (\textit{B = -0.480, 95\% CI [-0.87, -0.09], p = .02}).

\noindent\textbf{Mediation Effect on Perceived Fairness.}
No significant total effect on perceived fairness was found (\(p_{\text{total}} = .07\)). However, a significant indirect effect through SPP was observed (\textit{B = -0.302, 95\% CI [-0.55, -0.08], p = .02}), indicating a mediation effect, although the direct effect was not significant (\(p_{\text{direct}} = .0.8\)).

\noindent\textbf{Mediation Effect on Privacy and Emotional Response.}
The indirect effect was significant (\textit{B = 0.174, 95\% CI [0.05, 0.36], p = .02}) suggesting the presence of a mediation effect of SPP on PER. However, neither total nor direct effect was significant (\(p_{\text{total}} = .8\); \(p_{\text{direct}} = .3\));

\noindent\textbf{Effect on Impression Management.}
The total effect was significant (\textit{B = -0.27, 95\% CI [-0.52, -0.02], p = .03}) as was indirect effect (\textit{B = -0.19, 95\% CI [-0.37, -0.05], p = .02}). But the direct effect was not significant \(p_{\text{direct}} = .4\) suggesting the presence of a mediation effect through SPP.

\begin{framed}
\noindent
\textbf{\textit{Summary}}: For {white female} and {white male participants}, lower perceptions of the AI avatar's perceived social attributes (SPP) led to reduced fairness perceptions but improved privacy perceptions of the interview process. They also reported increased use of impression management (IM) tactics. For {black male participants}, higher SPP scores enhanced fairness perceptions and IM tactics but also raised privacy concerns. For {black female participants}, the mediation effect of SPP did not result in significant changes in their perception.
\end{framed}
\section{Discussion}

{In this study, we sought to better understand how demographic features such as the gender and race of an AVI agent could influence the interviewees' \textit{Perceived Fairness} (PF), \textit{Social Perception and Presence} (SPP), \textit{Privacy and Emotional Response} (PER), \textit{Impression Management} (IM). Additionally, we explored how the demographic attributes of different \textbf{interviewees} impact the perception of an AVI agent. These variables capture a nuanced view of the complex environment of an AVI interview where both social interactions and technological implications dictate an interviewee's experience. We also conducted a mediation analysis to explore how factors influencing interview experiences (e.g. PF, IM, PER) are mediated by the social presence and perception (SPP) of the AVI agents. In response to our RQ1, we discovered that the demographic attributes of the AVI interviewer agents did not markedly affect the participants' perceptions and responses. This contrasts with findings in the social psychology field, particularly in face-to-face (F2F) interviews with human interviewers, where differences are significant and attributes like stereotype threats are prominent. Conversely, in addressing our RQ2, we found that perceptions of PF, SPP, and IM varied notably across different participant demographics. 
Expanding on our findings for RQ2 we consistently observed significant differences in perceptions between black female and white female participants as well as between black male participants and white male participants regarding their evaluations of AVI agents (see Figure~\ref{fig:u_demo_direct_sig}).
This pattern suggests that the demographic factors of the \textbf{interviewee} were a key determining factor. However, no significant differences were observed in PER across any demographic categories. These findings underscore the complex interplay of personal identities in shaping user experiences with such AI technologies. The pronounced differences among participant groups, especially across gender and racial lines, emphasize the importance of considering diverse user perspectives in AI system design and implementation to foster equitable and inclusive user experiences.}

\subsection{Mediation Effect of Participant Demographics}

Through our mediation analysis, we identified SPP as a mediator in the relationship between participant demographics and their perceptions of AI-mediated interviews. Corroborating past findings \cite{Oh_Bailenson_Welch_2018} our analysis elucidates the nuances that lie across different interviewee demographics. This is evident in Figure~\ref{fig:spp_pd} where we can notice that the user's race was a key indicator of whether the participant had a heightened perception of the AVI agent's Social Perception and Presence (SPP). 

Overall, we find that SPP has a positive mediation effect on PF (see Figure~\ref{fig:direct_indirect}(a)) and IM (see Figure~\ref{fig:direct_indirect}(c)). A particularly interesting finding is the interaction between SPP and PER (Figure~\ref{fig:direct_indirect}(b)), marked by varying mediation effects. Although there was a negative correlation between SPP and PER, the impact varied across demographics: White female participants, without the mediation of SPP, exhibited lower PER scores, which reversed when the mediation effect of SPP was included. For black male participants, the opposite pattern was observed. 
These observations suggest that customizing AI systems to enhance social presence needs careful consideration of demographic-specific needs and expectations. This can significantly impact the interviewee's privacy perceptions and emotional response.

 \subsection{Design Implications}

{The differential impact of PF, SPP, PER, and IM across various demographic groups suggests 
the need for AVI systems to incorporate adaptive features. Customization should extend beyond visual representation to include elements like personality traits~\cite{li2017confiding} and conversational style~\cite{qiu2020estimating, qiu2020ticktalkturk}, which influence perceived social attributes such as SPP. By tailoring these aspects based on the interviewee's demographic factors, AVI systems can better align with the preferences of different users, 
thereby enhancing the interview experience. Given the heightened importance of perceived social attributes, we recommend design strategies that may adapt to 
demographic-specific conversational cues that acknowledge the user's background~\cite{Sue_2013}. This approach could make AI interviews more engaging and feel less transactional, promoting a positive interview experience. However, while designs that amplify social presence may be more effective for groups that respond positively to such cues, a different approach may be needed for those who find high social presence counterproductive or associate it with a sense of eeriness.}

\subsection{Caveats, Limitations, and Future Work}

{Our study focused on the screening phase of the recruitment process, a stage typically aimed at narrowing down the applicant pool. The dynamics and participant perceptions could differ significantly during the later stages of the hiring process, which merit further exploration as concomitant technologies continue to progress. Additionally, it is important to differentiate real interviews from our crowd-sourced experiment. In real interviews, the incentive is often to secure a job offer, whereas, in our study, participants may have not been driven by the same motivation \cite{draws2021checklist}.
}

{One of our main findings indicates that user demographic factors significantly affect perceptions of the AVI experience. Therefore, future studies could explore how adaptive and customizable features can enhance these perceptions.}

{Additionally, future research can delve deeper into how exactly the social perception and presence (SPP) positively or negatively influence perceptions of fairness and the ease of adopting impression management techniques, building on our initial findings. }

Lastly, while we present a unique outlook of AI-mediated interviews through our study we note some important ethical considerations while implementing such technology in practice. Our work takes the first strides towards advancing the understanding of how AI agents are perceived as a result of their race and gender in AVIs, to better understand how we can reduce biases in AVI screening processes. 
However, depending exclusively on AVI agents for interviews may result in candidates experiencing dehumanization or feeling evaluated solely through algorithmic assessments.
We believe that work in this realm should be embedded in a critical reflection of how human experiences in interview processes can be augmented and improved and not simply replaced. 
\section{Conclusions}

{The impact of the interviewer and the interviewee's race and gender in traditional recruitment interview processes has long been investigated in efforts to promote equity. With the rapid adoption of asynchronous video interview (AVI) processes, it is crucial to examine how these demographic factors influence outcomes in this new setting to inform design decisions. } 
Our findings from a $3\times2$ between-subjects factorial study indicate that the gender and race of AVI agents did not markedly affect interviewees' perceptions. However, notable differences emerged among various user demographic groups regarding their Perceived Fairness (PF) in the interview process, Social Perception and Presence (SPP) of the AVI agent, and utilization of Impression Management (IM) tactics. We discovered that Social Presence and Perception (SPP) mediates the relationship between participant demographics and their perceptions of the AI-mediated interview process.  The effects of SPP across different demographics suggest that enhancing the social presence of AI interviewers requires careful consideration of the specific needs and expectations of various groups to positively impact interviewees' privacy perceptions and emotional responses.

\section*{Acknowledgements}
This work was supported in part by the GENIUS (Generative
Enhanced Next-Generation Intelligent Understanding
Systems) ICAI lab, the TU Delft AI Initiative, and SHL Labs.

\bibliography{references}

\end{document}